\documentclass{article}

\usepackage{arxiv}

\usepackage{latexsym}
\usepackage{amsmath}
\usepackage{amssymb}
\usepackage{psfrag}

\usepackage{mathtools}
\usepackage{mathrsfs}
\usepackage{enumitem}
\usepackage{xcolor}
\usepackage{xpatch}
\usepackage{xr}
\usepackage{url}

\usepackage[utf8]{inputenc} 
\usepackage[T1]{fontenc}    
\usepackage{hyperref}       
\usepackage{url}            
\usepackage{booktabs}       
\usepackage{amsfonts}       
\usepackage{nicefrac}       
\usepackage{microtype}      
\usepackage{lipsum}		
\usepackage{graphicx}
\usepackage{natbib}
\usepackage{doi}

\usepackage{amsthm}
\newtheorem{theorem}{Theorem}
\newtheorem{lemma}{Lemma}

\renewcommand{\eqref}[1]{(\ref{#1})}

\let\longto\longrightarrow

\DeclareMathOperator{\var}{var}

\DeclareMathOperator{\G}{\mathcal{G}}

\DeclareMathOperator{\U}{\mathcal{U}}

\title{Nonparametric simulation extrapolation for measurement error models}

\author{ Dylan~Spicker \\
	Department of Epidemiology, Biostatistics, and Occupational Health\\
	McGill University\\
	Montreal, Canada \\
	\texttt{dylan.spicker@mcgill.ca} \\
	\And
	Michael P.~Wallace \\
	Department of Statistics and Actuarial Science\\
	University of Waterloo\\
	Waterloo, Canada \\
	\And
    Grace Y.~Yi \\ 
    Department of Statistical and Actuarial Sciences \\
    Department of Computer Science \\
    Western University \\
    London, Canada
}

\renewcommand{\shorttitle}{Nonparametric simulation extrapolation for measurement error models}

\hypersetup{
pdftitle={Nonparametric simulation extrapolation for measurement error models},
pdfauthor={Dylan~Spicker, Michael P.~Wallace, Grace Y.~Yi}
}

\begin{document}
\maketitle

\begin{abstract}
	The presence of measurement error is a widespread issue which, when ignored, can render the results of an analysis unreliable. Numerous corrections for the effects of measurement error have been proposed and studied, often under the assumption of a normally distributed, additive measurement error model. One such method is simulation extrapolation, or SIMEX. In many situations observed data are non-symmetric, heavy-tailed, or otherwise highly non-normal. In these settings, correction techniques relying on the assumption of normality are undesirable. We propose an extension to the simulation extrapolation method which is nonparametric in the sense that no specific distributional assumptions are required on the error terms. The technique is implemented when either validation data or replicate measurements are available, and is designed to be immediately accessible for those familiar with simulation extrapolation.
\end{abstract}


\section{Introduction}
Measurement error, where a variate of interest is not accurately observed, is a pervasive issue which can undermine the validity of an analysis. Numerous methods exist which correct for the effects of measurement error. These techniques commonly assume an additive model with normally distributed errors. This assumption, though appealing in its simplicity, is often unreasonable in real-world applications. See, for instance, \citet{GPS_Example, Income_Example, GeneExpression_Example, GalacticRotation_Example, BMI_Example, NRC_1986}, and \citet{Nusser_1996}. When the assumptions made are difficult to test, such as those regarding the distribution of error terms, concerns regarding the validity of correction procedures are amplified. 

The accommodation of non-normal errors is therefore an important area of study. Nonparametric and semiparametric methods, which do not impose strict assumptions on the distribution of the error terms, are important in providing flexible ways to correct for the effects of measurement error; see, for instance, \citet{BMI_Example,NP_Example1,NP_Example2,NP_Example3,CarrollBook}, and \citet{YiBook}. In addition to these nonparametric and semiparametric methods, several parametric techniques have been developed to account for non-normal errors \citep{Laplace,CoxNG}.

In order to facilitate error correction in the presence of non-normal errors, we present an extension of the commonly used simulation extrapolation method \citep{SIMEX}. Our extension does not assume that errors are normally distributed. Instead, we propose a nonparametric simulation extrapolation procedure, which consistently corrects for the effects of additive measurement error, regardless of the distribution of the error. The simulation extrapolation method is commonly abbreviated to SIMEX. We will refer to the standard SIMEX procedure as the P-SIMEX, or the parametric SIMEX, to distinguish it from our non-parametric procedure, which we refer to as the NP-SIMEX. It has been shown that the P-SIMEX method may be resilient to deviations from the assumption of normality of errors in some settings \citep{SIMEX}. Other authors have shown that when errors are non-normal the bias resulting from the P-SIMEX correction can be substantial \citep{Yi, Laplace}. 

The P-SIMEX is a three-step procedure, consisting of a simulation step, an extrapolation step, and an estimation step. The simulation step of the P-SIMEX has been described as a \emph{remeasurement method} \citep{MCCS}, emphasizing its similarities to bootstrap procedures. Just as bootstrap procedures can be made nonparametric by resampling from the empirical distribution, the P-SIMEX can be made nonparametric by \emph{remeasuring} using the empirical error distribution. Doing so allows for the relaxation of assumptions regarding the distribution of the error terms. We make explicit this nonparametric remeasurement procedure. This allows the NP-SIMEX to accommodate a wide range of error models, without making specific distributional assumptions, distinguishing it from the parametric extensions to simulation extrapolation for Laplace errors presented by \citet{Laplace}. 

\section{Background}
\subsection{Notation and Available Data}
Suppose that we have a sample indexed with $i\in\{1,\dots,n\}$. We take $Y_i$ to represent an outcome of interest, $X_i$ to be the explanatory variable which is subject to measurement error, and $Z_i$ to be the explanatory variables which are measured without error. The explanatory variable $X_i$ is taken to be univariate for ease of notation. We are concerned with estimating some parameter $\theta$ which relates the distribution of $Y_i$ to $X_i$ and $Z_i$. Instead of observing $X_i$ we observe $X_{i}^* = X_i + U_{i}$, where $U_{i}$ is the error-term, which we assume to be independent of $Y_i$, $X_i$, and $Z_i$. We assume that, in the error-free setting, we have an estimator $\widehat{\theta}(\{Y_i, X_i, Z_i\}_{i=1}^n)$ which is consistent for $\theta$.

Generally, measurement error correction techniques rely on auxiliary data to infer information about the errors. \emph{Validation data}, which can either be \emph{internal} or \emph{external}, involve the observation of the true $X_i$ alongside the proxy measurement $X_i^*$ for some set of individuals. Internal validation data refer to the situation where a subset of the sample of interest has this true measurement taken. That is, $\{Y_i, X_i, X_i^*, Z_i\}$ are observed for some subset of the $n$ total observations, while for the remaining only $\{Y_i, X_i^*, Z_i\}$ are observed. External validation data refer to the setting where, inside of the sample of interest, we have $\{Y_i, X_i^*, Z_i\}$ measured but pairs of $\{X_i, X_i^*\}$ are observed in a separate dataset. In order to make use of external validation data we must make assumptions regarding the \emph{transportability} of the error models. We assume that the error process for the external sample is equivalent to the error process within the sample of interest. This assumption allows us to use information from the external sample. 

When validation samples are not available we may instead use \emph{replicate measurements} together with some assumptions about the model. Here we observe repeated proxy measurements, $\{Y_i, X_{i1}^*, \dots, X_{i\kappa}^*, Z_i\}$ for all $i$, where, for $j=\{1,\dots,\kappa\}$, \begin{equation}X_{ij}^* = X_i + U_{ij},\label{eq::mem}\end{equation} and $\kappa$ is the number of replicates. It is further assumed that $U_{i1},\dots,U_{i\kappa}$ are independent and identically distributed according to $U_i$, and each $U_{ij}$ is independent of $\{X_i, Z_i, Y_i\}$. These auxiliary data can be used to estimate the variance of $U_i$, denoted $\sigma_U^2$. 

\subsection{Parametric Simulation Extrapolation}
We begin by briefly presenting the P-SIMEX. For full details of this technique see \citet{SIMEX} and \citet{SIMEX_Asymptotics}. For the P-SIMEX we assume that the error term $U_i$ is normally distributed, with mean zero and constant variance. We also assume that auxiliary data exist which allow for an estimate, $\widehat{\sigma}_U^2$, of $\sigma_U^2$. Once $\sigma_U^2$ has been estimated, the P-SIMEX then fixes a non-negative, real-valued grid with $M$ elements, denoted $\Lambda_\text{P}$, where $M$ is a user-specified positive integer. For each $\lambda_{\text{P}} \in \Lambda_\text{P}$, and each $i\in\{1,\dots,n\}$, we define the quantity $X_{bi}^*(\lambda_{\text{P}}) = \overline{X}_i^* + (\lambda_{\text{P}}^{1/2}\widehat{\sigma}_U)\nu_{bi}$, where $\nu_{bi} \sim \mathcal{N}(0,1)$ independent of all other terms, generated by the analyst, and $\overline{X}_i^*$ is the sample average of $X_{ij}^*$. This is considered to be a remeasured data set, which is then used to compute $\widehat{\theta}_{b}(\{Y_i,X_{bi}^*(\lambda_{\text{P}}), Z_i\}_{i=1}^n)$ by replacing $X_i$ with $X_{bi}^*(\lambda_{\text{P}})$ in the estimator obtained from the standard estimation method for error-free contexts. This process is repeated for $b\in\{1,\dots,B\}$ with $B$ specified by the analyst, producing a set of estimates which are then averaged as \[\widehat{\theta}(\{Y_i,X_i^*(\lambda_{\text{P}}), Z_i\}_{i=1}^n) = B^{-1}\sum_{b=1}^B \widehat{\theta}_b(\{Y_i,X_{bi}^*(\lambda_{\text{P}}), Z_i\}_{i=1}^n).\] After repeating this process across the entire grid the analyst will have a set of estimates and the corresponding $\lambda_\text{P}$ values, represented as $\{(\lambda_{\text{P}}, \widehat{\theta}(\{Y_i,X_i^*(\lambda_{\text{P}}),Z_i\}_{i=1}^n) : \lambda_\text{P} \in \Lambda_\text{P}\}$. These pairs of values can be used to fit a parametric relationship, $\G(\lambda)$, through least squares estimation, giving $\widehat{\G}(\lambda_\text{P})$, which describes $\widehat{\theta}$ as a function of the value of $\lambda_\text{P}$. Intuitively, $X_{bi}^*(\lambda)$ is such that $E[X_{bi}^*(\lambda)|X_i] = X_i$ and $\var(X_{bi}^*(\lambda)|X_i) = (\lambda+1)\sigma_U^2$. With $\lambda=-1$, $X_{bi}^*$ behaves as though it is $X_i$. The P-SIMEX estimator is taken to be $\widehat{\theta}_\text{P-SIMEX} = \widehat{\G}(-1)$. \citet{SIMEX} initiate this algorithm and \citet{SIMEX_Asymptotics} demonstrate that, under regularity conditions, including the availability of $\G(\cdot)$, $\widehat{\theta}_\text{P-SIMEX}$ is a consistent estimator of $\theta$.

If the measurement error is not normally distributed, or if $\G(\cdot)$ is not correctly specified, this method will produce only approximately consistent estimators with non-zero asymptotic bias. \citet{Laplace} provide a framework that can be adapted to different parametric error distributions by changing the distribution of $\nu_{bi}$ and allowing $\lambda_\text{P}$ to impact parameters other than the variance. These methods provide a flexible way to accommodate errors when there is a scientific rationale for specifying the distribution. Our proposal circumvents this requirement when no such rationale exists.

\section{Nonparametric Simulation Extrapolation}
\subsection{Estimation Procedure}
The key property of the P-SIMEX is that the characteristic function of $U_i + \left[\lambda\sigma_U^2\right]^{1/2}\nu_{bi}$ tends to $1$ as $\lambda$ tends to $-1$ \citep{Laplace}. We can exploit this property nonparametrically. Suppose that we are able to form the set $\U$, which contains all observed error terms for all individuals. Sampling $U^*$ from $\U$ is then sampling from the empirical distribution for the errors. Denote the characteristic function and distribution function of a random variable $X$ as $\varphi_{X}$ and $F_X$, respectively. Empirical distributions and characteristic functions are specified as $\widehat{F}_{X}$ and $\widehat{\varphi}_X$, respectively. That is, $\widehat{F}_X(x) = n^{-1}[I(X_1 \leq x)+\cdots+I(X_n \leq x)]$, where $I(\cdot)$ is an indicator function. 

Note that $U^*$ drawn from $\U$ has distribution function $\widehat{F}_U$ and characteristic function $\widehat{\varphi}_U(t)$. Denote the cardinality of a set $S$ as $|S|$, and take $|\U| = m$. Then as $m\to\infty$, $\widehat{\varphi}_U(t)\longto\varphi_U(t)$, for any $t\in\mathbb{R}$. Extending this, the characteristic function of $U + U^*$ converges pointwise to $\varphi_{U}^2$, as $m\to\infty$. Sampling $U_1^*, U_2^*,\dots,U_\lambda^*$ from $\U$, where $\lambda$ is a positive integer, the characteristic function of $U + U_1^* + \cdots U_\lambda^*$ converges pointwise to $\varphi_{U}^{\lambda + 1}$, as $m\to\infty$, since, for independent random variables $X$ and $W$, the sum $X + W$ will have a characteristic function $\varphi_{X}\varphi_{W}$. This function converges to $1$ as we take $\lambda$ to $-1$. This suggests the following as our proposed NP-SIMEX procedure. \begin{enumerate}
    \item Form the set $\U$. 
    \item Specify a fixed grid of non-negative integers, $\Lambda$. 
    \item For each $\lambda \in \Lambda$, $b\in\{1,\dots,B\}$, and every $i$, form \[\widetilde{X}_{bi}^*(\lambda) = X_i^* + \sum_{j=1}^{\lambda} U_{bi,j}^*,\] where the $U_{bi,j}^*$ are sampled independently, with replacement, from $\U$.
    \item Using $\widetilde{X}_{bi}^*(\lambda)$, compute $\widehat{\theta}_b(\{Y_i,X_{bi}^*(\lambda), Z_i\})$ for $b\in\{1,\dots,B\}$. Then compute \[\widehat{\theta}(\{Y_i,\widetilde{X}_i^*(\lambda), Z_i\}_{i=1}^n) = B^{-1}\sum_{b=1}^B \widehat{\theta}_b\left(\{Y_i,\widetilde{X}_{bi}^*(\lambda), Z_i\}\right).\]
    \item Fit a parametric regression model to $\{(\lambda, \widehat{\theta}(\{Y_i,\widetilde{X}_i^*(\lambda), Z_i\}_{i=1}^n)) : \lambda \in \Lambda\}$ and then extrapolate to $\lambda = -1$.
\end{enumerate}

This procedure relies on being able to form the set $\U$. The method for doing this depends on the auxiliary data that are available.

\subsection{Empirical Error Distribution Formation}
If we observe an internal validation sample, such that for $i\in\{1,\dots,n_1\}$ we observe $\{Y_i, X_i, X_i^*, Z_i\}$, and for $i=n_1+1, \dots, n$ we observe $\{Y_i, X_i^*, Z_i\}$, then we have implicitly observed a subsample including $U_i$. For $i\in\{1,\dots,n_1\}$ we can define $U_i = X_i^* - X_i$. Supposing then that the additive measurement error model is correct, we can form $\U$ as $\{U_i\}_{i=1}^{n_1}$. If we have an external validation sample, such that we have $\{Y_i, X_i^*, Z_i\}$ observed for $i\in\{1,\dots,n\}$, and $\{X_i, X_i^*\}$ observed in a separate dataset for $i\in\{1,\dots,n_1\}$, we can perform the same process and form $\U$ as the set of $X_i^* - X_i$ for $i\in\{1,\dots,n_1\}$. We require the same assumptions as with an internal validation sample, as well as the transportability assumption for external validation data. In either case no further restrictions are required on the distribution of $U$.

With replicate measurements we form $\U$ by restricting the error distributions that we consider to only those that are symmetric around a known constant. Given  that often observed errors follow heavy-tailed t-distributions, this assumption may be defensible \citep{Many_Examples, GalacticRotation_Example}.
We take the known constant to be $0$, without loss of generality. Assume that for all $i\in\{1,\dots,n\}$ we observe $\{Y_i, X_{i1}^*, \dots, X_{ik}^*, Z_i\}$, where $X_{ij}^*$ are as in equation (\ref{eq::mem}). Take $k=2$ as an example. Then, consider $\widetilde{X}_i^* = 0.5\left(X_{i1}^* + X_{i2}^*\right)$. By equation (\ref{eq::mem}), $\widetilde{X}_i^* = X_i + 0.5\left(U_{i1} + U_{i2}\right)$, which can as such be viewed as an error-prone measurement of $X_i$. If we define $\widetilde{U}_i = 0.5\left(X_{i1}^* - X_{i2}^*\right)$, then by symmetry we have that $\widetilde{U}_i = 0.5\left(U_{i1} - U_{i2}\right) \stackrel{d}{=} 0.5\left(U_{i1} + U_{i2}\right)$, where $X \stackrel{d}{=} W$ means that $X$ and $W$ are equal in distribution. Following from this, $\widetilde{X}_i^* \stackrel{d}{=} X_i + \widetilde{U}_i$. As a result, we can form $\U = \{\widetilde{U}_{1}, \dots, \widetilde{U}_{n}\}$, which serves as the set to sample from when using the mean response. This procedure also can be applied when $k \neq 2$ (see the details in the Appendix). 

\subsection{Illustration and Theoretical Justification}\label{subsec::theory}
We first illustrate our procedure with a simplified example. Suppose that we wish to estimate the fourth moment of $X$ (assuming its existence) using $\widehat{\theta}(X) = n^{-1}(X_1^4 + \cdots + X_n^4)$. In place of $X_i$ we observe $X_{ij}^* = X_i + U_{ij}$, for $j=1,2$, where the $U_{ij}$ are symmetrically distributed about $0$, and independent of $X_i$ and each other. Taking the mean observation for each individual as $\widetilde{X}_i^*$, then an application of the weak law of large numbers demonstrates that as $n\to\infty$, we have $\widehat{\theta}(\{\widetilde{X}_i^*\}_{i=1}^n) \stackrel{p}{\longto} \mu_4 + 6\mu_2\sigma_{U}^2 + E[\widetilde{U}^4]$, where $\mu_j = E[X^j]$ for $j=2,4$. This is generally biased for $\mu_4$, rendering the naive estimator inconsistent.

To apply the P-SIMEX consider $\widehat{\theta}(\{X_{bi}^*(\lambda_{\text{P}})\}_{i=1}^n)$, which has a limit in probability as $n\to\infty$, of $\mu_4 + 6\mu_2(1+\lambda_{\text{P}})\sigma_U^2 + E[(\widetilde{U} + \nu^*)^4]$, with $\nu^* \sim \mathcal{N}(0,\lambda_\text{P}\sigma_U^2)$. This can be expanded to $\mu_4 + 6\mu_2(1+\lambda_{\text{P}})\sigma_U^2 + 3\lambda_{\text{P}}^2\sigma_{U}^4 + 6\lambda_{\text{P}}\sigma_U^4+ E[\widetilde{U}^4]$. Since $E[\widetilde{U}^4]$ is functionally independent of $\lambda_{\text{P}}$, we can take $\G(\lambda) = a + b\lambda + c\lambda^2$ to be the extrapolant. Then $\G(-1) = \mu_4 + E[\widetilde{U}^4] - 3\sigma_U^4$. When $U_{ij} \sim \mathcal{N}(0, \sigma_U^2)$, then $E[\widetilde{U}^4] = 3\sigma_U^4$, and the P-SIMEX procedure consistently corrects for the effects of measurement error. If instead we take $U_{ij} \sim t_{5}$ then $E[\widetilde{U}^4] = 25$. Combined with $\sigma_{U}^2 = 25/3$ the P-SIMEX procedure leaves a residual asymptotic bias of $50/3$.

Consider applying the NP-SIMEX, with $\widehat{\theta}(\{X_i + \widetilde{U}_{i}(\lambda)\}_{i=1}^n)$, where $\widetilde{U}_{i}(\lambda) \stackrel{d}{=}\widetilde{U}_0 + \cdots + \widetilde{U}_\lambda$. Making the same argument as above gives $\mu_4 + 6\mu_2(1 + \lambda)\sigma_U^2 + (\lambda + 1)E[\widetilde{U}^4] + 3(\lambda + 1)\lambda\sigma_U^4$. This produces an extrapolant that is exactly quadratic, giving $\G'(\lambda) = a' + b'\lambda + c'\lambda^2$. This leads to the conclusion that $\G'(-1) = \mu_4$, regardless of the value of $E[\widetilde{U}^4]$. The analytic tractability of this example allows it to serve as an illustration of the conditions required for consistency. 

In this example we treated the NP-SIMEX technique as though the estimator was computed based on random observations that are distributed as $X + \widetilde{U}_0 + \cdots + \widetilde{U}_\lambda$. In practice, we will compute the estimator based on random quantities distributed as $X + U + U_1^* + \cdots + U_\lambda^*$, where the $U_j^*$ are sampled from $\mathcal{U}$. We argued above that as $|\U|\to\infty$, this quantity will have characteristic function $\varphi_X\varphi_{U}^{\lambda+1}$, and as such will behave as though it were distributed as $X + \widetilde{U}_0 + \cdots + \widetilde{U}_\lambda$. In general, this requires smoothness assumptions on the estimator. Take $F_\lambda$ to be the distribution function for the sum of $X$ and $\lambda+1$ copies of $U$, given by the convolution $F_X\ast F_U^{\ast(\lambda+1)}$ and assume that $m=n$. Then we establish the following asymptotic results.

\begin{theorem}{Theorem 1}{}
     Suppose that both $X$ and $U$ are absolutely continuous with respect to the Lebesgue measure, the estimator from the error-free context can be expressed as a functional over the distributions $\mathbf{T}$, and that $\mathbf{T}(F_\lambda) = \mathcal{G}(\lambda)$ for all $\lambda \geq -1$, where $\mathcal{G}(\lambda)$ has a known parametric form. If $\mathbf{T}$ is continuous or bounded with respect to $\Vert\cdot\Vert_\infty$, then $\widehat{\theta}_\text{NP-SIMEX}$ is consistent for $\theta$, as  $n\to\infty$.
\end{theorem}
\begin{theorem}{Theorem 2}{}
     Suppose that both $X$ and $U$ are absolutely continuous with respect to the Lebesgue measure, the estimator from the error-free context can be expressed as a functional over the distributions $\mathbf{T}$, and that $\mathbf{T}(F_\lambda) = \mathcal{G}(\lambda)$ for all $\lambda \geq -1$, where $\mathcal{G}(\lambda)$ has a known parametric form. If $\mathbf{T}$ is Fréchet differentiable, with respect to $\Vert\cdot\Vert_\infty$, then $\sqrt{n}(\widehat{\theta}_\text{NP-SIMEX} - \theta)$ has an asymptotic normal distribution (as $n\to\infty$) with mean $0$.
\end{theorem}

These consistency and asymptotic distributional results are established through non-standard, technical asymptotic theory. This makes the conditions for application of these results difficult to assess in practice. These results rely on treating the estimator from the error-free context, $\widehat{\theta}$, as a functional over distributions, and then assessing the continuity or differentiability of these functionals. We establish sufficient conditions for consistency which are comparatively straightforward to check, though the condition of Fréchet differentiability of the statistical functional that is sufficient for asymptotic normality is fairly strong, and difficult to translate into the standard language of estimators. \citet{Shao1993} demonstrates that large classes of commonly used estimators satisfy this condition (for instance, differentiable functions of the mean, large classes of M-estimators, and the Cramér-von Mises test statistic). Other authors have noted that there are many statistical functionals of interest for which Fréchet differentiability with respect to $\Vert\cdot\Vert_\infty$ is not satisfied, but where consistency results can still be obtained \citep{Regularity4}.

We expect that, using more sophisticated arguments, asymptotic normality could be obtained under weaker forms of differentiability. However, it is worth considering the other strong assumption that is being made: that the extrapolant is known and correctly specified. This is an assumption shared by other SIMEX estimators \citep{SIMEX,Laplace} but it is quite strong nonetheless. In practice, this assumption is the reason that SIMEX is often treated as an approximately consistent technique for the correction of the effects of measurement error, where the use of a suitable extrapolant can reduce bias, even if it does not entirely eliminate it \citep{SIMEX,CarrollBook}.

Variance estimation can be conducted through a bootstrap procedure. In certain settings this may be undesirable due to a need for nested resampling procedures. With the P-SIMEX two additional variance estimation techniques were proposed: one using a modified SIMEX procedure \citep{SIMEX_Jackknife} and one using the asymptotic distribution \citep{SIMEX_Asymptotics}. Theorem~2 allows for the use of sandwich estimation techniques to establish an estimate of the asymptotic variance. The details are provided in full in \citet{SIMEX_Asymptotics}. To do this we require an estimate for the covariance of the stacked influence curves of the functional representation of the estimator. In settings where these representations are common, this result can be useful. For other situations, where both bootstrap and the asymptotic distribution are not viable, the techniques proposed by \citet{SIMEX_Jackknife} can be adopted for the NP-SIMEX; this is outlined in the Appendix.

\subsection{Extensions of the Core Procedure}
The presentation of the NP-SIMEX procedure assumed that, when using replicate data to form $\mathcal{U}$, these replicates were independent and identically distributed. This assumption is stronger than is necessary. First, define a contrast vector $a$ such that $0 = a_1 + \cdots + a_k$ and $1 = |a_1| + \cdots + |a_k|$. Suppose that $U_{ij}$ are symmetric and independent, but not identically distributed. For each $j$, $a_jU_{ij} \stackrel{d}{=} |a_j|U_{ij}$. Then, to form elements of the set $\U$, we can take \[U_i^* = \sum_{j=1}^k a_jX_{ij}^* = \sum_{j=1}^ka_jU_{ij} \stackrel{d}{=} \sum_{j=1}^k|a_j|U_{ij}.\] Here $U_i^*$ represents a realization from the empirical error distribution for \[\widetilde{X}_i^* = \sum_{j=1}^k |a_j|X_{ij}^* = X_i + \sum_{j=1}^k|a_j|U_{ij}.\] 

We can slightly relax the assumption of symmetric errors with replicate data. Assuming that $U_{ij}$ is symmetric, for any $a_j$, we have $a_jU_{ij} \stackrel{d}{=} |a_j|U_{ij}$. When $a_j \geq 0$, then $a_jU_{ij} = |a_j|U_{ij}$. Supposing that $U_{ij}$ does not follow a symmetric distribution, then if $a_j \geq 0$ the previous argument still holds. As a result, so long as at least one of the repeated measurements has an error distribution which is symmetric, the NP-SIMEX can proceed by defining the contrast vector $a$ as before, with the restriction that each non-symmetric entry has a positive value.

Finally, we have made the common assumption that $U_{ij}\perp X_i$, for all $i,j$. It may be the case that errors depend on the true, underlying value, rendering the presented argument for the NP-SIMEX invalid. If $U_{ij}$ are dependent on $X_i$, the errors associated with individual $i$ will be drawn from a different distribution than those from individual $i' \neq i$. Conceptually, we can replace the empirical distribution with a distribution estimated using kernel density estimation (KDE). There have been many proposed techniques for estimating a conditional density based on kernel methods \citep{Hall_2004}. With an estimated conditional KDE, denoted $\widehat{f}_{U|X}$, it is possible to sample directly from this conditional distribution (see for instance \citet[ Section~14.7]{KDE_Book}). The details of this technique are expanded upon in the Appendix. 

While this procedure conceptually works, the difficulty is that we cannot directly condition on $X_i$ outside of the validation sample. Instead, we need a method for drawing from the correct error distribution, given only $X_i^*$. A possible technique is to repeat this procedure, using the validation sample to estimate $\widehat{f}_{X|X^*}(x|x^*)$, and then draw $\widetilde{X}_i$ based on this KDE, for each individual in the sample. This could then be used as the value of $X_i$ to condition on. An alternative approach is to draw error realizations directly from the distribution of $\widehat{f}_{U|X^*}(u|x^*)$. This procedure can be directly applied over the complete sample. Despite the easier application, this procedure only approximately corrects for dependence in the errors, even in the limit, as generally conditioning on $X^*$ induces dependence between $X$ and $U$, even where none previously existed.

The discussion of using KDEs in the event that $U_i\not\perp X_i$ may also prompt consideration of using KDEs under the assumption that $U_i\perp X_i$. Instead of forming $\U$ directly, we can estimate $\widehat{f}_U(u)$, and then sample from this KDE. This procedure is the \emph{smoothed bootstrap} \citep{SmoothedBootstrap}. In certain settings smoothing can improve the performance of estimators, particularly with small sample sizes \citep{Efron_1982}. This smoothing could be applied, under the independence assumption, with either validation or repeated measurements. We refer to this as the smoothed NP-SIMEX.

\section{Simulation Studies}
In this section we present four simulation studies investigating the behaviour of the estimator in several scenarios. Additional simulation results are provided in the Appendix. The first simulation contrasts the P-SIMEX and the NP-SIMEX in a logistic regression. We take $n=5000$, $B=100$, and a $\Lambda$ grid size of $10$. We generate a true, unobserved variable $X$ according to a $\mathcal{N}(1, 2)$ distribution, and consider the outcome such that $P(Y = 1 | X) = H(1 - X)$, where $H(\cdot)$ is taken as the inverse-logit function. In place of $X$, we generate two replicated responses for each individual, $X_1$ and $X_2$, which are given by $X + U_j$, $j=1,2$ where $U_j$ follows a t-distribution, independent of all other variables, with degrees of freedom in $\{3, 4, 5, 10, 30\}$. Both the P-SIMEX and NP-SIMEX are implemented using the nonlinear extrapolant, $\mathcal{G}(\lambda) = a + b/(c + \lambda)$, and we compute $95\%$ confidence intervals using a bias adjusted bootstrap procedure with $500$ bootstrap replicates. These simulations are repeated $200$ times and the results are shown in Table~\ref{tab::simulation_1}, where the columns under the heading MSE report the mean squared error over the $200$ repeated simulations, the mean bias reports the average bias over the $200$ simulations, and the columns for the coverage probability report the proportion of constructed $95\%$ bootstrap confidence intervals which contain the true value.

\begin{table}
    \caption{\label{tab::simulation_1} The mean squared error (MSE), mean bias, and coverage probability from 200 replicate simulations, estimating the slope parameter in a logistic regression, where the variate has t-distributed error, with varying degrees of freedom (DFs) presented. Coverage probability is computed using a bias corrected bootstrap, with 500 bootstrap resamples.}
    
    \centering
    \begin{tabular}{rrrrrlrrr}
        \hline
        & \multicolumn{3}{c}{P-SIMEX} && \multicolumn{3}{c}{NP-SIMEX} \\
        \cline{2-4}\cline{6-8}
        DFs & MSE & Mean Bias & Coverage && MSE & Mean Bias & Coverage \\
        \hline
        3 & 0.014 & -0.108 & 0.200 && 0.002 & -0.005 & 0.920 \\
        4 & 0.014 & -0.111 & 0.165 && 0.001 & 0.002 & 0.930 \\
        5 & 0.015 & -0.116 & 0.110 && 0.001 & 0.005 & 0.930 \\
        10 & 0.016 & -0.119 & 0.095 && 0.001 & 0.002 & 0.940 \\
        30 & 0.016 & -0.120 & 0.100 && 0.001 & 0.001 & 0.950 \\
       \hline
    \end{tabular}
\end{table}

Across all t-distributions the NP-SIMEX dramatically improves over the P-SIMEX in MSE and bias. The computed coverage probabilities are also substantially improved, though there is evidence of under coverage, particularly for low degrees of freedom. While none of these differences are significant at a 95\% level, these anti-conservative results warrant caution and careful application of the bootstrap procedure, specifically when the error distribution is likely to be particularly heavy-tailed. The results suggest that bootstrapping may be feasible for quantifying the uncertainty in the NP-SIMEX procedure, when the computation is not a problem.

The second simulation investigates the impact of sample size on the variability of the estimation. We use the example from Section~\ref{subsec::theory}, which involves estimating the fourth moment of $X$ from a $\mathcal{N}(5, 4)$ distribution. We take two error-prone measurements, both subject to additive error from a $t_5$ distribution. The errors are independent of each other, and of the $X$ variables. We vary the sample size from $100$ to $100 000$, replicating each $1000$ times. We take $M=10$ and $B=500$. The MSE over the $1000$ replicates when the truth is available, the relative MSEs (the observed MSE divided by the observed MSE when the truth is used), and the mean bias for the naive, P-SIMEX, the NP-SIMEX, and the smoothed NP-SIMEX corrections are shown in Table~\ref{tab::simulation_2}.
\begin{table}
    \caption{\label{tab::simulation_2}The relative MSE (and mean bias) from 1000 replicated simulations, estimating the fourth moment of a contaminated random variable, over different sample sizes ($n$). Main values are the MSE divided by the MSE computed using the error-free variable (truth), at the same sample size; bracketed values are the mean biases. The MSE and bias using the true values are given under the heading ``Truth''.}
    \centering
    \begin{tabular}{rrrrrr}
        \hline
        $n$ & Naive & P-SIMEX & NP-SIMEX & Smoothed NP & Truth \\
        \hline
        100 & 2.279 (148.305) & 1.682 (9.600) & 1.273 (-12.185) & 1.299 (-23.709) & 34647.104 (-6.013) \\
        500 & 5.048 (151.618) & 1.771 (34.205) & 1.386 (3.955) & 1.431 (-2.972) & 6660.699 (2.029) \\
        1000 & 8.275 (148.976) & 1.759 (35.874) & 1.383 (-1.329) & 1.390 (-6.045) & 3350.250 (0.672) \\
        5000 & 34.404 (147.637) & 3.561 (36.599) & 1.456 (-1.495) & 1.467 (-4.127) & 664.606 (-1.443) \\
        10000 & 70.330 (149.476) & 5.259 (35.557) & 1.364 (0.751) & 1.349 (-1.344) & 324.617 (0.483) \\
        20000 & 133.467 (149.920) & 8.622 (34.759) & 1.464 (0.068) & 1.484 (-1.383) & 170.587 (0.134) \\
        50000 & 350.646 (149.509) & 21.772 (35.821) & 1.446 (0.491) & 1.540 (-0.053) & 64.119 (0.320) \\
        100000 & 683.371 (149.037) & 39.205 (35.045) & 1.547 (-0.164) & 1.564 (-0.442) & 32.579 (0.032) \\
        \hline
    \end{tabular}
\end{table}

Predictably, the naive method performs unsatisfactorily, and demonstrates the utility of both the P-SIMEX and the NP-SIMEX in reducing the impacts of measurement error. While the MSE is quite large for small $n$ regardless of the method, this is also true for the true estimator, seeing only an 8.3\% and 11.3\% increase in the relative MSEs over truth for the NP-SIMEX and the P-SIMEX respectively (when $n=100$). While the raw MSE decreases for both correction procedures as $n$ increases, the relative MSE increases for both. However, the NP-SIMEX remains relatively comparable to the truth for all values of $n$, while for larger values of $n$, the P-SIMEX performs substantially worse. The P-SIMEX and the naive estimator are left with substantial bias, even for large sample sizes, where the NP-SIMEX appears to mostly eliminate it. There is no substantial difference between the smoothed and unsmoothed estimators.

Our third simulation considers the use of validation data in place of replicate measurements. We generate the true variate, $X$, to be Gamma with shape parameter $1$ and scale parameter $2$, such that $E(X) = 2$ and $\var(X) = 4$. We then contaminate it with an additive error term, $U_i$, which is mean-zero and follows a Laplace distribution. We consider several values for the measurement error variance, taking the ratio $\sigma_U/\sigma_X$ to be one of $0.1$, $0.5$, $1$, or $2$. The sample size is selected to be one of $\{1000, 10000, 100000\}$, and we assume that an internal validation sample is available comprsised of $5\%$, $10\%$, or $50\%$ of the total sample. All results use the nonlinear extrapolant for both the P-SIMEX and the NP-SIMEX. The MSEs for the naive, P-SIMEX, and NP-SIMEX estimators across all scenarios are presented in Tables~\ref{tab::simulation_3} and \ref{tab::simulation_3B}.
\begin{table}
    \caption{\label{tab::simulation_3}The MSE from 1000 replicated simulations estimating the slope parameter in a logistic regression, over different sample sizes ($n$), validation sample size percentages (\%), and ratios of standard deviations ($\sigma_U/\sigma_X$). The results compare the naive estimators, those from the parametric SIMEX (P), and those from the nonparametric SIMEX (NP). This table contains results with a sufficiently large validation sample, relative to the measurement error variance.}
    \centering
    \begin{tabular}{lccclccclccclccc}
      \hline
        & \multicolumn{3}{c}{$\sigma_U/\sigma_X = 0.1$} && \multicolumn{3}{c}{$\sigma_U/\sigma_X = 0.5$} && \multicolumn{3}{c}{$\sigma_U/\sigma_X = 1$}  && \multicolumn{3}{c}{$\sigma_U/\sigma_X = 2$} \\
        \cline{2-4}\cline{6-8}\cline{10-12}\cline{14-16}
        \% & Naive & P & NP && Naive & P & NP && Naive & P & NP && Naive & P & NP \\
        \hline
        \multicolumn{16}{c}{$n=1000$}\\
        5 & .011 & .011 & .011 && .225 & .114 & .124 && -- & -- & -- && -- & -- & -- \\
        10 & .011 & .011 & .012 && .220 & .108 & .039 && -- & -- & -- && -- & -- & -- \\
        50 & .012 & .012 & .013 && .221 & .108 & .024 && .726 & .264 & .086 && -- & -- & -- \\
        \hline
        \multicolumn{16}{c}{$n=10000$}\\
        5 & .002 & .002 & .001 && .219 & .103 & .009 && .726 & .259 & .071 && -- & -- & -- \\
        10 & .002 & .002 & .001 && .220 & .103 & .007 && .726 & .257 & .048 && -- & -- & -- \\
        50 & .002 & .002 & .001 && .219 & .102 & .006 && .724 & .254 & .035 && 1.230 & .856 & .092 \\
        \hline
        \multicolumn{16}{c}{$n=100000$}\\
        5 & .001 & .001 & .000 && .219 & .102 & .004 && .725 & .254 & .033 && 1.230 & .858 & .100 \\
        10 & .001 & .001 & .000 && .219 & .102 & .004 && .725 & .254 & .032 && 1.230 & .857 & .076 \\
        50 & .001 & .001 & .000 && .219 & .102 & .004 && .725 & .254 & .032 && 1.230 & .857 & .068 \\
       \hline
    \end{tabular}
\end{table}

In Table~\ref{tab::simulation_3} we see that the NP-SIMEX outperforms both the P-SIMEX procedure and naive estimation, particularly when the ratio of variances grows. With a small validation sample, and with small measurement error, we see that the P-SIMEX procedure performs at the same level as the NP-SIMEX. However, as the estimators stabilize, by increasing either $n$ or the proportion of validation samples, the NP-SIMEX correction substantially outperforms either of the other methods. This table excludes results with a ratio of standard deviations equal to $1$ when the validation sample is $100$ samples or fewer, and the results where the ratio of standard deviations was $2$ for validation samples up to $1000$ individuals. The results of these scenarios are provided in Table~\ref{tab::simulation_3B}.
\begin{table}
    \caption{\label{tab::simulation_3B}The MSE from 1000 replicated simulations estimating the slope parameter in a logistic regression, over different sample sizes ($n$), validation sample size percentages (\%), and ratios of standard deviations ($\sigma_U/\sigma_X$). The results compare the naive estimators, those from the parametric SIMEX (P), and those from the nonparametric SIMEX (NP). This table contains results with an insufficiently large validation sample, relative to the measurement error variance.}
    \centering
    \begin{tabular}{lccclccc}
      \hline
        & \multicolumn{3}{c}{$\sigma_U/\sigma_X = 1$}  && \multicolumn{3}{c}{$\sigma_U/\sigma_X = 2$} \\
        \cline{2-4} \cline{6-8}
        $n$ (\%) & Naive & P & NP && Naive & P & NP \\
        \hline
        $1000$ (5) & 0.7280 & 0.2945 & 9917.3388 && 1.2303 & 0.8716 & 52.9248 \\
        $1000$ (10) & 0.7239 & 0.2733 & 100.5551 && 1.2277 & 0.8601 & 190.2874 \\
        $1000$ (50) & -- & -- & -- && 1.2310 & 0.8627 & 190.6604 \\
        $10000$ (5) & -- & -- & -- && 1.2302 & 0.8593 & 124.5868 \\
        $10000$ (10) & -- & -- & -- && 1.2305 & 0.8593 & 58.1180 \\
       \hline
    \end{tabular}
\end{table}

These results demonstrate the instability of the nonparametric procedure at small sample sizes, when the error is sufficiently large. Note that, as would be expected, the naive estimators are not impacted by the size of the validation sample, and the impact on the P-SIMEX is fairly small. The results demonstrate that if the validation sample is too small or the estimated variation too large, nonparametric techniques are not appropriate. This emphasizes the importance of considering the fact that, when using validation data, convergence of the correction happens in $n_1$ rather than $n$. Fortunately, while these results demonstrate clear instability at small sample sizes, the breakdown in performance is easy to see. The techniques, when unstable, will often result in estimates which are unreasonable from a subject matter perspective. We stress careful application of these techniques in settings where sample sizes may lead to instability.

In the final simulation we demonstrate the viability of the NP-SIMEX strategy using kernel density estimation. We take $X \sim \mathcal{N}(1, 4)$, and $U|X \sim \mathcal{N}(\rho(X - 1), 1)$, where $\rho$ is a parameter selected from $\{0, 0.5, 1, 2\}$, and a binary outcome, with $P(Y = 1|X) = H(1 - X)$. The sample size is $n=1000$ with a $20\%$ validation sample, and the simulations are repeated $500$ times. Within this context, we compare four different estimation strategies: first, a standard application of SIMEX; second, a version of the NP-SIMEX where we first sample $X$ from $X|X^*$, and then from $U|X$; third, a version of the NP-SIMEX where we sample directly from $U|X^*$; and finally, sampling directly $X|X^*$, and averaging over many iterations of this. The results of the MSE for the slope parameter estimate are contained in Table~\ref{tb::conditional_sim_results}.

\begin{table}[ht]
    \centering
    \caption[Logistic regression slope estimates using the KDE-modified SIMEX methods for dependent errors.]{MSE of logistic regression slope parameter estimates from 500 simulation runs where there is simulated dependence between the true variate ($X$) and the error term ($U$) and the strength of this relationship is mediated by $\rho$.}
    \begin{tabular}{ccccc}
        \hline
         $\rho$ & SIMEX & NP-SIMEX ($U|X$) & NP-SIMEX ($U|X^*$) & Direct from $X|X^*$ \\
         \hline 
           0 & 0.012 & 0.014 & 0.053 & 0.183 \\
         0.5 & 0.067 & 0.010 & 0.012 & 0.070 \\
         1   & 0.115 & 0.013 & 0.009 & 0.034 \\
         2   & 0.222 & 0.018 & 0.009 & 0.011 \\ \hline
    \end{tabular}
    \label{tb::conditional_sim_results}
\end{table}

When there is independence between $U$ and $X$, the standard SIMEX estimators perform well. As this dependence strengthens, the corrections are less able to address concerns due to measurement error. The NP-SIMEX which fits directly to $U|X$ sees relatively comparable performance across all of the scenarios tested, and is always among the best techniques. Drawing from $U|X^*$ performs comparatively poorly when there is independence, but with dependent errors it sees a marked improvement, performing better than any other technique. The averaging of samples directly from $X|X^*$ tends to perform fairly well again with a strong dependence, though it seems unlikely to be preferable to the NP-SIMEX using $U|X^*$ or $U|X$. 

\section{Data Analysis}
We now consider an analysis of the Korean Longitudinal Study of Aging (KLoSA), following that conducted by \citet{BMI_Example}. The KLoSA considers South Korean citizens, aged 45 and over, in a longitudinal study looking to determine health effects of aging. Our analysis considers data on $n=9842$ individuals, with an internal validation sample of $n_1=505$, and we are interested in estimating how an individual's BMI impacts their propensity towards being hypertensive. In the main study BMI is estimated through self-reported weight and heights, and the validation sample includes clinical measurements, taken to be the truth, alongside the self-reported values. We also observe each individual's age, which we consider to be error-free. 

\begin{figure}
    \centering
    \makebox{\includegraphics[width=0.65\textwidth]{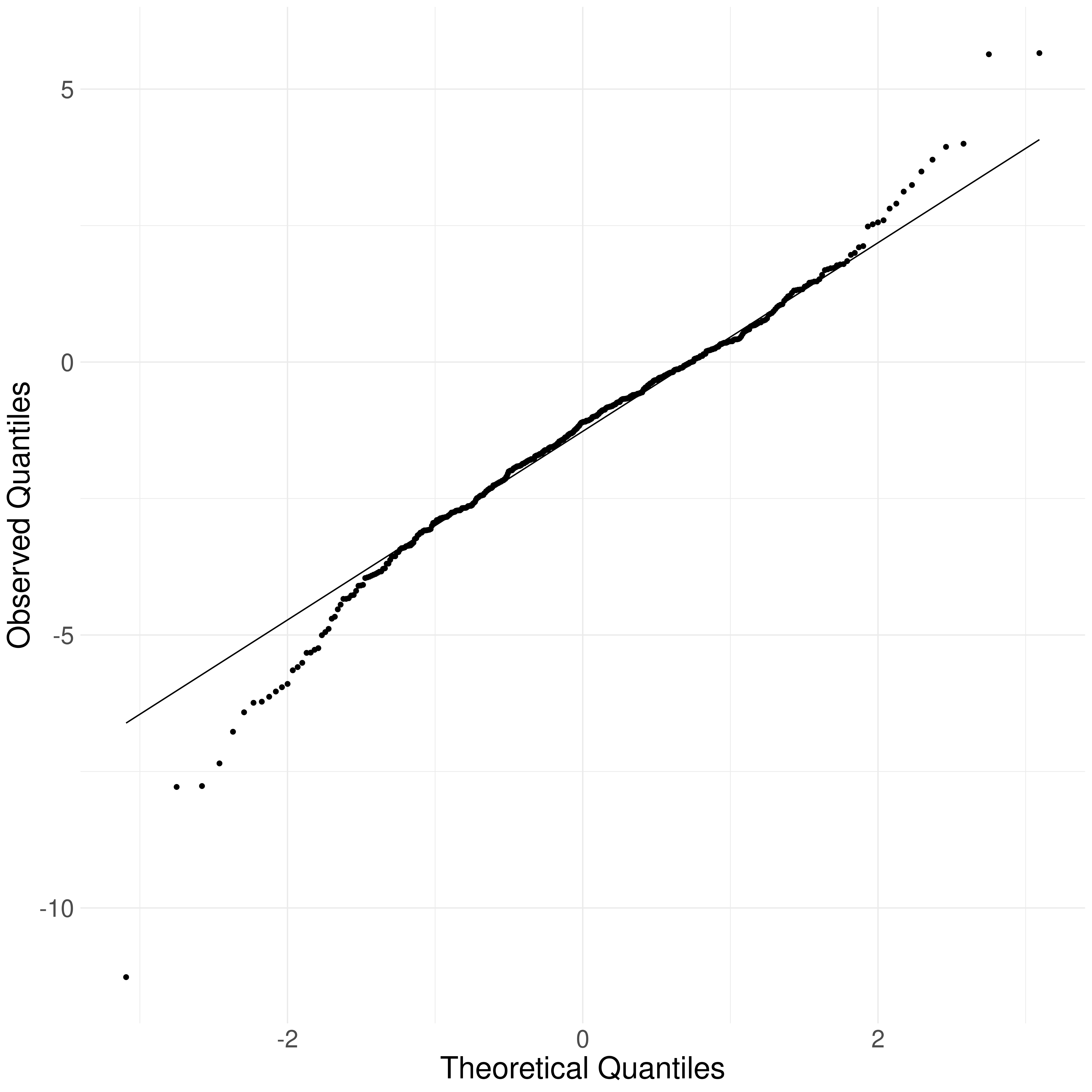}}
    \caption[KLoSA Normal Q-Q plot.]{\label{fig::klosa_qq}A normal Q-Q plot for the observed errors in the Korean Longitudinal Study of Aging.} 
\end{figure}

An analysis of the validation sample demonstrates that the errors are non-normal, as is evidenced by the Q-Q plot in Figure~\ref{fig::klosa_qq}, an excess kurtosis of $2.05$, and negative skew. This suggests that the standard P-SIMEX procedure may not be appropriate. We estimate the ratio of $\sigma_U/\sigma_X$, using the $505$ validation sample observations, as $0.898$.

We analyze these data fitting a logistic regression model, with a logit link function, including the main effects of BMI and age. That is, we assume that \[\text{logit}(E[Y_i|X_i, Z_i]) = \beta_0 + \beta_1\text{BMI}_i + \beta_2\text{Age}_i.\] We generate bootstrap standard error estimates with $1000$ replicates, and compare both the NP-SIMEX and the P-SIMEX, using the nonlinear extrapolant. We also consider an uncorrected analysis. The results are summarized in Table~\ref{tab::klosa_analysis}.

\begin{table}
    \centering
    \caption[Slope parameter estimates and standard errors for an analysis of KLoSA using SIMEX correction techniques.]{Point estimates and bootstrap standard error (SE) estimates for the logistic regression parameters, estimating the propensity of hypertension with an intercept $(\beta_0)$, self-reported BMI ($\beta_1$), and age ($\beta_2$). The estimates are based on $1000$ bootstrap replicates, comparing the naive method, P-SIMEX correction, and NP-SIMEX correction.}
    \begin{tabular}{lcclcclcc}
        \hline
        & \multicolumn{2}{c}{$\beta_0$} && \multicolumn{2}{c}{$\beta_1$} && \multicolumn{2}{c}{$\beta_2$} \\
        \cline{2-3} \cline{5-6} \cline{8-9}
        Method & Estimate & SE && Estimate & SE && Estimate & SE \\ \hline
        Naive & -5.023 & 0.439 && 0.030 & 0.016 && 0.053 & 0.002 \\        
        P-SIMEX & -5.512 & 0.833 && 0.049 & 0.031 && 0.054 & 0.003 \\        
        NP-SIMEX & -5.061 & 0.673 && 0.039 & 0.026 && 0.054 & 0.002 \\ \hline
    \end{tabular}
    \label{tab::klosa_analysis}
\end{table}

The three methods tend to agree on the estimate and standard error for $\beta_2$. For both $\beta_0$ and $\beta_1$, we see that the NP-SIMEX method estimates values which are larger in magnitude than the naive estimator but smaller than the P-SIMEX correction, both for the point estimate and the standard error. All three techniques suggest a positive effect of BMI on hypertension, though, the level of significance of this effect varies dramatically: $0.054$, $0.117$, and $0.133$ for the naive, P-SIMEX, and NP-SIMEX estimators respectively.

\begin{figure}
    \centering
    \includegraphics[width=0.65\textwidth]{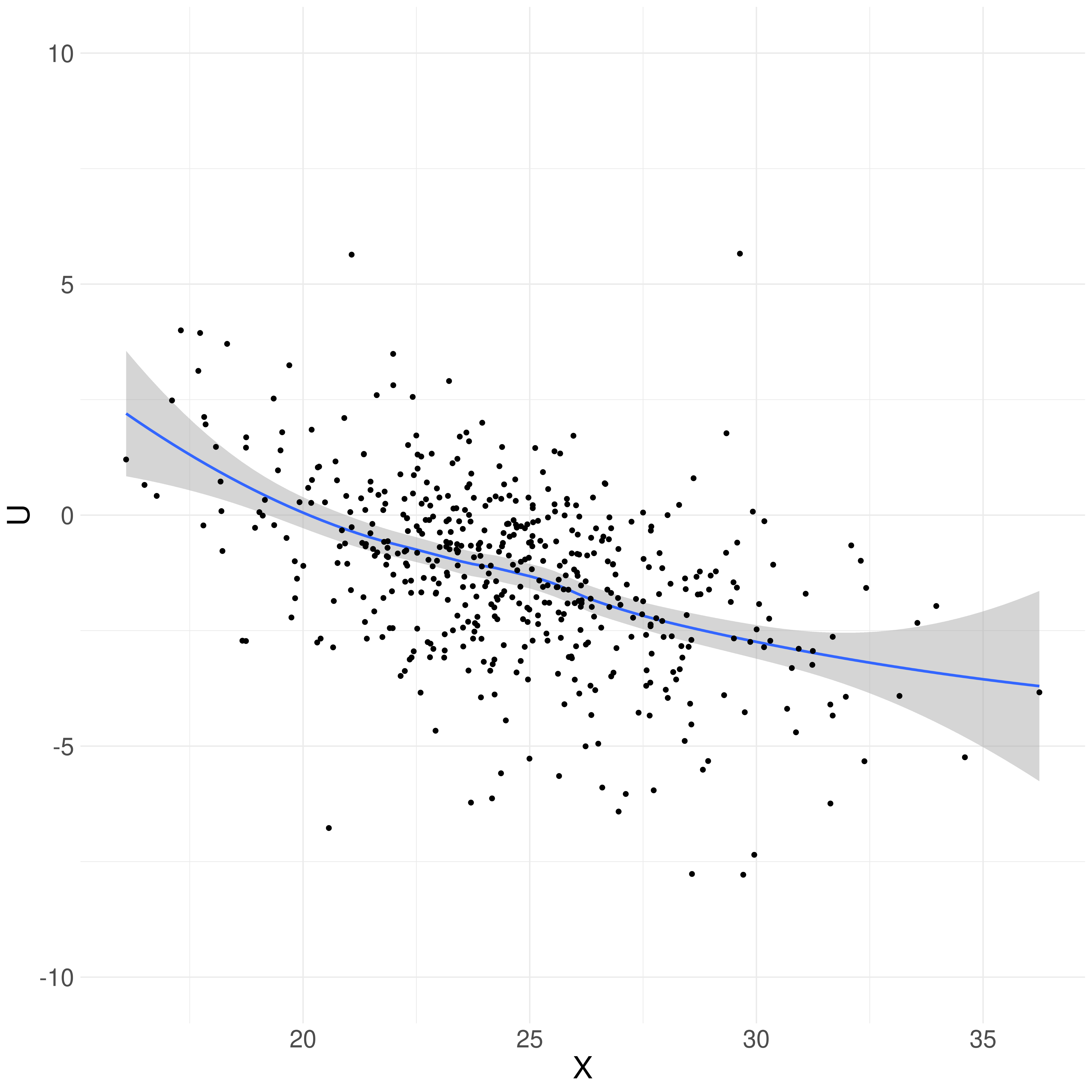}
    \caption[KLoSA observed errors versus truth in the validation sample.]{Estimated errors ($U$) versus the true underlying BMI for individuals within the KLoSA validation sample. The included line is a LOESS curve, included to clearly delineate the degree of dependence observed within these data.}
    \label{fig::dependent_errors_KLoSA}
\end{figure}

One concern with this analysis of KLoSA is that there is evidence that the observed errors are not independent of the true values. In Figure~\ref{fig::dependent_errors_KLoSA} we can see a plot of the error terms versus the true values, illustrating the degree of dependence that is present in these data. This relationship corresponds to a correlation of approximately $-0.464$. This pattern has also been observed in past research \citep{Villanueva_2001}.

We consider conducting the same analysis using the two proposed KDE NP-SIMEX estimation techniques, based on both sampling first from $X|X^*$ and then $U|X$, and on sampling directly from $U|X^*$. The estimated coefficients and bootstrap standard errors are included in Table~\ref{tb::klosa_analysis_conditional}. For each of the analyses we consider using both the quadratic and the nonlinear extrapolant. The slope coefficient for age ($\beta_2$) generally was not estimable with the nonlinear extrapolant and so these results are not reported.

\begin{table}
    \centering
    \caption[Slope parameter estimates and standard errors for an analysis of KLoSA using the KDE-modified SIMEX correction techniques.]{Point estimates and bootstrap standard error (SE) estimates for the logistic regression parameters, estimating the propensity of hypertension with an intercept $(\beta_0)$, self-reported BMI ($\beta_1$), and age ($\beta_2$). The estimates are based on $500$ bootstrap replicates, comparing the conditional NP-SIMEX method using $U|X$, and the conditional NP-SIMEX method using $U|X^*$, both with a quadratic and nonlinear extrapolant.}
    \begin{tabular}{llcclcclcc}
        \hline
        & & \multicolumn{2}{c}{$\beta_0$} && \multicolumn{2}{c}{$\beta_1$} && \multicolumn{2}{c}{$\beta_2$} \\
        \cline{3-4}\cline{6-7}\cline{9-10}
        Method & Extrapolant & Estimate & SE && Estimate & SE && Estimate & SE \\
        \hline
        $U|X$ & Quadratic & -5.531 & 0.472 && 0.049 & 0.017 && 0.054 & 0.002 \\
        $U|X$ & Nonlinear & -4.740 & 0.669 && 0.063 & 0.028 && -- & --  \\
        $U|X^*$ & Quadratic & -5.515 & 0.473 && 0.049 & 0.017 && 0.054 & 0.002 \\
        $U|X^*$ & Nonlinear & -4.936 & 0.679 && 0.061 & 0.038 && --  & -- \\
       \hline
    \end{tabular}
    \label{tb::klosa_analysis_conditional}
\end{table}

The resulting estimates for $\beta_0$ and $\beta_2$ do not differ substantially from the non-conditional results. The signs for these coefficients, and their approximate magnitudes are comparable to the previously estimated values. The largest difference is in the estimates for $\beta_1$ when the nonlinear extrapolant was used. These results suggest that the magnitude of the effect size was severely underestimated. Comparing the use of the nonlinear extrapolant with either of the conditional distributions to that of the previous analyses we find that the previous estimates had magnitudes which were between $0.5$ and $0.8$ times the estimated magnitude using the conditional distribution. The p-values for a test of significance using $U|X$ and $U|X^*$ were respectively $0.022$ and $0.105$.

Given the clear dependence observed between the errors and the true BMI in these data, and in past literature, we advise taking the conditional analyses as more reliable for estimators of the truth than the unconditional analyses presented originally. Agreement on the intercept and age coefficient gives confidence in these estimates.

\section{Discussion}
Measurement error is a ubiquitous issue which impacts the validity of statistical inference. Methods for correcting for measurement error often rely on untestable assumptions regarding the distribution of the error terms. The simulation extrapolation procedure is a commonly used correction procedure which can correct for the effects of measurement error in a wide variety of models \citep{YiBook,CarrollBook}. Where the corrections are not consistent, simulation extrapolation can reduce the bias present in a naive analysis. While the P-SIMEX is easy to implement, it relies on an assumption that the errors are normally distributed \citep{SIMEX,SIMEX_Asymptotics,SIMEX_Jackknife}. We provide a nonparametric simulation extrapolation procedure, which relaxes this assumption, allowing for any error distribution if validation data are available, or any symmetric error distribution when replicate measurements are available. The implementation of our procedure is similar to the P-SIMEX, allowing for analysts familiar with the P-SIMEX to adopt this nonparametric extension where appropriate. Our simulation results demonstrate that, when there is sufficient data and the models used are conducive to simulation extrapolation, the proposed procedure effectively corrects for the effects of measurement error across a wide variety of measurement error models. These results are supported by large-sample, theoretical justifications. 

Our results complement those of \citet{Laplace}, who accommodate non-normal errors through parametric extensions to simulation extrapolation. These parametric extensions carry the standard benefits of parametric methods. We would expect them to be more efficient, particularly at small sample sizes, when the assumed error distribution is correct. In contrast, the NP-SIMEX provides a flexible way to accommodate errors without the need to specify a particular distribution for the error terms. This is an important extension for settings which may not have strong, subject matter justifications for making a particular distributional assumption.

While nonparametric corrections for the effects of measurement error are important, given the prevalence of non-normal error distributions, there are drawbacks to such an approach. Primarily, nonparametric methods tend to require substantially more data to behave in a stable manner, and our correction is no exception. 

All the code used to perform the simulations and analyses are available publicly on GitHub, at \url{https://github.com/DylanSpicker/np-simex}.

\subsection*{Acknowledgments}
This research was partially supported by funding from the Natural Sciences and Engineering Council of Canada (NSERC). Yi is a Canada Research Chair in Data Science (Tier 1). Her research was undertaken, in part, thanks to funding from the Canada Research Chairs Program. The authors thank the anonymous reviewers for their helpful comments.

\bibliographystyle{unsrtnat}
\bibliography{references}  

\pagebreak 

\appendix
\section*{Empirical Error Distribution with $k\neq2$ replicates}\label{appdx::contrasts}

In the paper we presented the technique for forming $\mathcal{U}$ when there are $k=2$ replicates. If we have $k\neq 2$ a similar process can be followed. Consider defining a $k$-dimensional contrast, $(a_1, \dots, a_k)$ such that $0 = a_1 + \cdots + a_k$ and $1 = |a_1| + \cdots + |a_k|$. Then, consider the two sums \[\sum_{j=1}^k a_jX_{ij}^* \quad \text{ and } \quad \sum_{j=1}^k |a_j|X_{ij}^*.\] Since $a_1 + \cdots + a_k = 0$ the former becomes \[\sum_{j=1}^k a_jU_{ij} \stackrel{d}{=} \sum_{j=1}^k |a_j|U_{ij},\] where the distributional equality holds since the $U_{ij}$ are symmetric by assumption. For the latter term, we note that since $|a_1| + \cdots + |a_k| = 1$, this simplifies to $X_i + |a_1|U_{i1} + \cdots + |a_k|U_{ik}$. As a result, we can take $\widetilde{X}_{i}^* = |a_1|X_{i1}^* + \cdots + |a_k|X_{ik}^*$, and $\widetilde{U}_{i} = a_1X_{i1}^* + \cdots + a_kX_{ik}^*$, and apply the same argument as above. In the case of $k=2$, we have used the contrast $(1/2, -1/2)^{\top}$. This is naturally extended to \[\left(\underbrace{1/k, \dots, 1/k}_{\text{$k/2$ terms}}, \underbrace{-1/k, \dots, -1/k}_{\text{$k/2$ terms}}\right)^{\top},\] for $k$ even, and when $k$ is odd, \[\left(\underbrace{1/(k+1), \dots, 1/(k+1)}_{\text{$(k+1)/2$ terms}}, \underbrace{-1/(k-1), \dots, -1/(k-1)}_{\text{$(k-1)/2$ terms}}\right)^{\top}.\]

\section*{Proof of Theorems}
In order to prove Theorems 1 and 2, we first present several Lemmas. Recall that $\widehat{F}_{n,m,\lambda}$ represents the empirical CDF of $X_{i}^*(\lambda)$ and $F_{m,\lambda} = F_X\ast F_U\ast \widehat{F}_U^{\ast\lambda}$ when $|\mathcal{U}| = m$, and $F_\lambda = F_X\ast F_U^{\ast(\lambda+1)}$. Moreover, we take $m=n$.
\begin{lemma}{Lemma A1.}{}
    Suppose that $X$ is a random variable which is absolutely continuous with respect to the Lebesgue measure, with density $f$ and distribution function $F$. Take $G$ to be a distribution function, and $G_n$ to be a sequence of distribution functions, then \begin{enumerate}
        \item $\Vert F\ast G_n - F\ast G\Vert_\infty = \Vert F\ast(G_n - G)\Vert_\infty \leq \Vert G_n - G\Vert_\infty$; and
        \item If $\Vert G_n - G\Vert \stackrel{a.s.}{\longto} 0$ as $n\to\infty$, then $\Vert F\ast(G_n - G)\Vert \stackrel{a.s.}{\longto} 0$ as $n\to\infty$.
    \end{enumerate}
\end{lemma}
\begin{proof}{}{Proof of Lemma A1}
    Note that this is essentially a specific case of Young's Convolution Inequality. Also note that (2) is immediate from (1), so we prove the inequality. Suppose that $f$ is a density function. \begin{align*}
        \Vert F\ast(G_n - G)\Vert_\infty &= \sup_{x\in\mathbb{R}}\left|\int_{-\infty}^\infty f(t)\left[G_n(x-t)-G(x-t)\right]dt\right| \\
        &\leq \sup_{x\in\mathbb{R}}\int_{-\infty}^\infty \left|f(t)\left[G_n(x-t)-G(x-t)\right]\right|dt \leq \Vert f \Vert_p\Vert G_n - G\Vert_q,
    \end{align*} where $1/p + 1/q = 1$, $p,q\in[1,\infty]$. The second inequality is simply Hölder's Inequality. Take $q = \infty$ so that $p=1$. Take $f$ to be a pmf instead. Then we have \begin{align*}
        \Vert F\ast(G_n - G)\Vert_\infty &= \sup_{x\in\mathbb{R}}\left|\sum_{t=-\infty}^\infty f(t)\left[G_n(x-t)-G(x-t)\right]\right| \\
        &\leq \sup_{x\in\mathbb{R}}\sum_{t=-\infty}^\infty \left|f(t)\right|\cdot\left|\left[G_n(x-t)-G(x-t)\right]\right| \\
        &\leq \sup_{x\in\mathbb{R}}\Vert G_n - G\Vert_\infty\sum_{t=-\infty}^\infty \left|f(t)\right| = \Vert G_n - G\Vert_\infty.
    \end{align*}
\end{proof}

\begin{lemma}{Lemma 1.}{}
    Suppose both $X$ and $U$ are absolutely continuous with respect to the Lebesgue measure. Then, for every $\lambda\in\Lambda$, as $n\to\infty$, \[\sup_{x\in\mathbb{R}}|\widehat{F}_{n,m,\lambda}(x) - F_\lambda(x)| = \Vert\widehat{F}_{n,m,\lambda} - F_\lambda\Vert_{\infty} \stackrel{a.s.}{\longto} 0.\] 
\end{lemma}

\begin{proof}{}{Proof of Lemma 1}
    First note that, by the triangle inequality we have that \[\Vert\widehat{F}_{n,m,\lambda} - F_\lambda\Vert_{\infty} \leq \Vert\widehat{F}_{n,m,\lambda} - F_{m,\lambda}\Vert_\infty + \Vert F_{m,\lambda} - F_\lambda\Vert_\infty.\] Appealing to \citet[Theorem 2.1]{Shorack1979} we get, as $n\to\infty$ \[\Vert\widehat{F}_{n,m,\lambda} - F_{m,\lambda}\Vert_\infty \stackrel{a.s.}{\longto} 0,\] since $m=n$ and for every fixed $m$, the distribution of the realizations is given by $F_{m,\lambda}$. 
    
    The second term converges almost surely to zero as well, which we show inductively in $\lambda$. Take $\lambda = 0$. Then $F_{m,\lambda} = F_\lambda = F_X\ast F_U$, and so $\Vert F_{m,\lambda} - F_\lambda\Vert_\infty = 0$. Consider $\lambda = 1$. Then, $F_{m,\lambda} = F_\lambda = F_X\ast F_U\ast\widehat{F}_U$ and $F_\lambda = F_X\ast F_U\ast F_U$, and we can apply Lemma A1 with $F = F_X\ast F_U$, $G_n = \widehat{F}_U$, and $G = F_U$. Assume that for a positive integer $\lambda'$, this result holds for all $\lambda = 2,3,\dots,\lambda'-1$. 
    
    Note that we can write $\widehat{F}_{n,m,\lambda} = \widehat{F}_U\ast F_{n,m,\lambda-1}$ and $F_{\lambda} = F_{\lambda-1}\ast F_U$. From these identities and the triangle inequality, with $\lambda=\lambda'$ we get \[\Vert \widehat{F}_{n,m,\lambda'} - F_{\lambda'}\Vert_\infty \leq \Vert\widehat{F}_U\ast\widehat{F}_{n,m,\lambda'-1}-\widehat{F}_U\ast F_{\lambda'-1}\Vert_\infty + \Vert F_{\lambda'-1}\ast\widehat{F}_U - F_{\lambda'-1}\ast F_U\Vert_\infty.\] The first term converges almost surely to $0$ by the inductive hypothesis and Lemma A1. The second term converges almost surely to $0$ by the standard Glivenko-Cantelli Theorem, and Lemma A1.
    
\end{proof}

\begin{lemma}{Lemma 2(i).}{}
    Suppose that the conditions of Lemma 1 are satisfied, and that the estimator from the error-free context can be expressed as a functional over distributions $\mathbf{T}(\cdot)$, which is continuous with respect to $\Vert\cdot\Vert_\infty$ at $F_\lambda$ for all $\lambda\in\Lambda$, then $\widehat{\theta} = \mathbf{T}(\widehat{F}_{n,m,\lambda})$ is consistent for $\mathbf{T}(F_\lambda)$, as $n\to\infty$.
\end{lemma}

\begin{proof}{}{Proof of Lemma 2(i)}
    By Lemma 1 we get that $\Vert \widehat{F}_{m,n,\lambda} - F_\lambda\Vert_\infty \stackrel{a.s.}{\longto} 0$. Continuity of the functional here means that, for all $\epsilon > 0$, there exists $\delta > 0$ such that $\Vert \widehat{F}_{n,m,\lambda} - F_\lambda\Vert_\infty < \delta \implies |\mathbf{T}(\widehat{F}_{n,m,\lambda}) - \mathbf{T}(F_\lambda)| \leq \epsilon$. As a result, for $\epsilon \geq 0$, we have \begin{align*}
        P\left[\Vert \widehat{F}_{n,m,\lambda} - F_\lambda\Vert_\infty < \delta\right] & \leq P\left[|\mathbf{T}(\widehat{F}_{n,m,\lambda}) - \mathbf{T}(F_\lambda)| \leq \epsilon\right] \\
        \implies \lim_{n\to\infty} P\left[\Vert \widehat{F}_{n,m,\lambda} - F_\lambda\Vert_\infty < \delta\right] &\leq \lim_{n\to\infty} P\left[|\mathbf{T}(\widehat{F}_{n,m,\lambda}) - \mathbf{T}(F_\lambda)| \leq \epsilon\right] \leq 1 \\
        \implies 1 &\leq \lim_{n\to\infty} P\left[|\mathbf{T}(\widehat{F}_{n,m,\lambda}) - \mathbf{T}(F_\lambda)| \leq \epsilon\right] \leq 1.
    \end{align*} Where the last line follows from $\Vert \widehat{F}_{m,n,\lambda} - F_\lambda\Vert_\infty \stackrel{a.s.}{\longto} 0$.
\end{proof}

\begin{lemma}{Lemma 2(ii).}{}
    Suppose that the conditions of Lemma 1 are satisfied, and that the estimator from the error-free context can be expressed as a functional over distributions $\mathbf{T}(\cdot)$, which is bounded with respect to $\Vert\cdot\Vert_\infty$, then $\widehat{\theta}_{\text{NP-SIMEX}} = \mathbf{T}(\widehat{F}_{n,m,\lambda})$ is consistent for $\mathbf{T}(F_\lambda)$, as $n\to\infty$.
\end{lemma}

\begin{proof}{}{Proof of Lemma 2(ii)}
    By boundedness we mean that, for a constant $C$, \[|\mathbf{T}(\widehat{F}_{n,m,\lambda}) - \mathbf{T}(F_\lambda)| \leq C\Vert\widehat{F}_{n,m,\lambda} - F_\lambda\Vert_\infty.\] As a result, for $\epsilon > 0$, \[P\left[|\mathbf{T}(\widehat{F}_{n,m,\lambda}) - \mathbf{T}(F_\lambda)| < \epsilon\right] \geq P\left[\Vert\widehat{F}_{n,m,\lambda} - F_\lambda\Vert_\infty < \frac{\epsilon}{C}\right].\] This inequality is the same as in the proof of Lemma 2A, taking $\delta = \epsilon/C$, giving the necessary result.
\end{proof}

\begin{lemma}{Lemma 3}{}
    Suppose that both $X$ and $U$ are absolutely continuous with respect to the Lebesgue measure. Then \[\sqrt{n}\Vert\widehat{F}_{m,n,\lambda} - \widehat{F}_\lambda\Vert_\infty = O_p(1),\] under regularity conditions (e.g., those in \citet[ Theorem 2.2]{Shorack1979}).
\end{lemma}
\begin{proof}{}{Proof of Lemma 3}
    Consider that \begin{align*}
        \sqrt{n}\Vert\widehat{F}_{m,n,\lambda} - F_\lambda\Vert_\infty &= \sqrt{n}\Vert\widehat{F}_{m,n,\lambda} - F_{m,\lambda} + F_{m,\lambda} - F_\lambda\Vert_\infty \\
        &\leq \sqrt{n}\Vert\widehat{F}_{m,n,\lambda} - F_{m,\lambda}\Vert_\infty + \sqrt{n}\Vert F_{m,\lambda} - F_\lambda\Vert_\infty \\
        &= O_p(1) + \sqrt{n}\Vert F_{m,\lambda} - F_\lambda\Vert_\infty.
    \end{align*} Here the first term is $O_p(1)$ by \citet[ Theorem 2.2]{Shorack1979}. As a result, we show that $\sqrt{n}\Vert F_{m,\lambda} - F_\lambda\Vert_\infty = O_p(1)$. 
    
    We will show that $\Vert F_{m,\lambda} - F_\lambda\Vert_\infty \leq C\Vert\widehat{F}_U - F_U\Vert_\infty$, which gives the necessary response. The argument is inductive in $\lambda$. For $\lambda = 0$ note that $F_{m,\lambda} = F_X\ast F_U$ and $F_\lambda = F_X\ast F_U$, so the conclusion holds trivially. Assume that, for $\lambda\in\{1,\dots,\lambda'-1\}$ our inductive hypothesis (IH) is given by \[\Vert F_{m,\lambda} - F_\lambda\Vert_\infty \leq C_\lambda\Vert\widehat{F}_U - F_U\Vert_\infty \tag{IH}.\] Then for $\lambda=\lambda'$ take \[\Vert F_{m,\lambda'} - F_{\lambda'}\Vert_\infty = \Vert F_{X}\ast F_U\ast\left(\widehat{F}_U^{\ast\lambda'} - F_U^{\ast\lambda'}\right)\Vert_\infty \leq \Vert\widehat{F}_U^{\ast\lambda'} - F_U^{\ast\lambda'}\Vert_\infty,\] where the inequality follows from Lemma A1. Splitting these terms this then gives us
\begin{align*}
        \Vert\widehat{F}_U^{\ast\lambda'} - F_U^{\ast\lambda'}\Vert_\infty &= \Vert\widehat{F}_U^{(\ast[\lambda'-1])}\ast\widehat{F}_U - \widehat{F}_U\ast F_U^{(\ast[\lambda'-1])} + \widehat{F}_U\ast F_U^{(\ast[\lambda'-1])} - F_U\ast F_U^{(\ast[\lambda'-1])}\Vert_\infty \\
        &= \Vert\widehat{F}_U\ast\left(\widehat{F}_U^{(\ast[\lambda'-1])} - F_U^{(\ast[\lambda'-1])}\right) + F_U^{(\ast[\lambda'-1])}\ast\left(\widehat{F}_U-F_U\right)\Vert_\infty.
\end{align*}
Using this relation, with the triangle inequality we then get that,
\begin{align*}
        \Vert\widehat{F}_U^{\ast\lambda'} - F_U^{\ast\lambda'}\Vert_\infty &\leq \Vert\widehat{F}_U\ast\left(\widehat{F}_U^{(\ast[\lambda'-1])} - F_U^{(\ast[\lambda'-1])}\right)\Vert_\infty + \Vert F_U^{(\ast[\lambda'-1])}\ast\left(\widehat{F}_U-F_U\right)\Vert_\infty \\
        &\leq \Vert\widehat{F}_U^{(\ast[\lambda'-1])} - F_U^{(\ast[\lambda'-1])}\Vert_\infty + \Vert\widehat{F}_U-F_U\Vert_\infty \\
        &\leq (C_{\lambda'-1} + 1)\Vert\widehat{F}_U - F_U\Vert_\infty,
    \end{align*} where the second inequality is from Lemma A1, and the last from the inductive hypothesis (IH). Then, since \[\sqrt{n}\Vert F_{m,\lambda} - F_\lambda\Vert_\infty \leq C_\lambda\sqrt{n}\Vert\widehat{F}_U - F_U\Vert_\infty = O_p(1),\] we have the necessary result.
\end{proof}

\begin{lemma}{Lemma 4.}{}
    Suppose that the conditions of Lemma 3 are satisfied, and that the estimator from the error-free context can be expressed as a functional over distributions $\mathbf{T}(\cdot)$, which is Fréchet differentiable with respect to $\Vert\cdot\Vert_\infty$ at $F_\lambda$. Then \[\sqrt{n}\left(\mathbf{T}(\widehat{F}_{n,m,\lambda}) - \mathbf{T}(F_\lambda)\right) \stackrel{d}{\longto} \mathcal{N}(0,E[\psi_{F_\lambda}^2(X)]),\] where $\psi_F$ is the influence function of $\mathbf{T}$ at $F$.
\end{lemma}

\begin{proof}{}{Proof of Lemma 4}
    This follows almost immediately from Fréchet differentiability and Lemma 3. By Fréchet differentiability we have \begin{align*}
        &\quad\sqrt{n}\left(\mathbf{T}(\widehat{F}_{n,m,\lambda}) - \mathbf{T}(F_\lambda)\right) \\
        &= \sqrt{n}\int\psi_{F_\lambda}d\widehat{F}_{n,m,\lambda} + \sqrt{n}o(\Vert\widehat{F}_{n,m,\lambda}-F_\lambda\Vert_\infty) \\
        &= \sqrt{n}\int\psi_{F_\lambda}d\widehat{F}_{n,m,\lambda} + \sqrt{n}\Vert\widehat{F}_{n,m,\lambda}-F_\lambda\Vert_\infty\frac{o(\Vert\widehat{F}_{n,m,\lambda}-F_\lambda\Vert_\infty)}{\Vert\widehat{F}_{n,m,\lambda}-F_\lambda\Vert_\infty} \\
        &= \sqrt{n}\int\psi_{F_\lambda}d\widehat{F}_{n,m,\lambda} + O_p(1)o(1).
    \end{align*} Now, as $n\to\infty$, \[\sqrt{n}\int\psi_{F_\lambda}d\widehat{F}_{n,m,\lambda} = \frac{1}{\sqrt{n}}\sum_{i=1}^n\psi_F(X_{i}^*(\lambda)) \stackrel{d}{\longto} \mathcal{N}(0,E[\psi_{F_\lambda}^2(X)]),\] which combines with Slutsky's theorem to give the result.
\end{proof}

\begin{proof}{}{Proof of Theorems 1 and 2}
    The proofs for Theorems 1 and 2 follow directly from the arguments in \citet{SIMEX_Asymptotics} in addition to Lemmas 2(i) and 2(ii) (for consistency), and Lemma 4 for asymptotic normality. In particular, note that $\mathbf{T}(\lambda)$ can be replaced by $\mathcal{G}(\lambda)$ in the statement of the Lemmas, which gives consistency (asymptotic normality) of $\mathbf{T}(\widehat{F}_{n,m,\lambda})$ to the extrapolant. Then, by assuming that the extrapolant is correctly specified, consistency follows through extrapolation. For asymptotic normality, the argument progresses as-in \citet[ Section 3.3]{SIMEX_Asymptotics}.
\end{proof}

\section*{NP-SIMEX with Kernel Density Estimation}
In order to implement the KDE NP-SIMEX procedure, we need to consider the conditional density of $U_i|X_i$. We can view the sample, within the validation set, as observing $\{Y_i, X_i, U_i, Z_i\}$ for each individual (or $\{X_i, U_i\}$ in the event of an external validation set). Then, we can take \begin{align*}
    \widehat{f}_{U|X}(u|x) &= \frac{\widehat{f}_{U,X}(u, x)}{\widehat{f}_{X}(x)}; \\
    \widehat{f}_{U,X}(u, x) &= \frac{1}{n_1h_Uh_X}\sum_{i=1}^{n_1}K_{U}\left(\frac{u - u_i}{h_U}\right)K_{X}\left(\frac{x - x_i}{h_X}\right); \\
    \widehat{f}_{X}(x) &= \frac{1}{n_1h_X}\sum_{i=1}^{n_1} K_{X}\left(\frac{x - x_i}{h_X}\right),
\end{align*} where $h_U$ and $h_X$ are bandwidth parameters (selected based on the observed data), and $K_U(\cdot)$ and $K_X(\cdot)$ are kernel functions (for instance, the Gaussian kernel). Estimation of the bandwidth parameters was addressed by \citet{Hall_2004}, where they use cross validation based on the integrated squared error. Once estimated, the bandwidth parameters can be used to sample from the conditional distribution, given a particular value of $X=x$. Specifically, to sample conditional on $X = x$, we select an individual $i\in\{1,\dots,n_1\}$ from the validation set weighted proportional to $K_X\left([x - x_i]/\widehat{h}_X\right)$. We then draw a realization from the $K_U$ distribution, based on the kernel parameter $\widehat{h}_U$, centred at $u_i$. Using Gaussian kernels, this results in drawing a random realization from a normal distribution with mean $u_i$ and variance $\widehat{h}_U^2$ \citep{KDE_Book}. The analysis conducted previously can then proceed, conditional on $X_i = x_i$. The convergence of this modified procedure will be in $n_1$ rather than $n$. If necessary we can also consider conditioning on additional factors (say $Z_i$) if those are strongly informative. 

For the smoothed NP-SIMEX procedure, instead of forming $\U$ directly, we can estimate $\widehat{f}_U(u)$, and then sample from the this KDE. To do so, with equal probability we sample an index $i \in \{1,\dots,n_1\}$ and then draw from the distribution corresponding to $K_U$ with bandwidth parameter $\widehat{h}_U$, centred at $u_i$. 

\section*{Additional Simulations}
These simulations extend the previous setting of estimating the fourth moment, this time assuming that there are three replicated observations. The errors are taken to be contaminated normals, as in the first set of simulations, with $\rho = 0$ and $\rho = 0.5$, and $X$ remains distributed as a $\mathcal{N}(5, 4)$ random variable. The sample size is fixed at $n=15000$, with $B=500$, and $M=10$. These simulations are replicated $1000$ times. These results are repeated with two available replicates. The results are shown in Table~\ref{tab::simulation_APP1}.
\begin{table}
    \caption{\label{tab::simulation_APP1}The MSE and median based on three replicates, estimating the fourth moment (true value $1273$) of a contaminated random variable, with either contaminated normal or normal errors. The results compare having two replicates available to the same estimators having three replicates available for the correction.}
    \centering
    \begin{tabular}{lrrlrr}
        \hline
        & \multicolumn{2}{c}{Two Replicates} && \multicolumn{2}{c}{Three Replicates} \\
        \cline{2-3}\cline{5-6}
        Error Distribution & MSE  & Median && MSE & Median \\
        Normal & 279.076 & 1273.261 && 270.212 & 1272.946\\
        Contaminated Normal ($\rho = 0.5$) & 4074.825 & 1269.713 && 2011.741  & 1268.731  \\
         \hline
      \end{tabular}
\end{table}
We can see a reduction in MSE for both error distributions, with a far more substantial improvement coming when the errors are drawn from a contaminated normal distribution. In this specific context the addition of a third replicate decreased the MSE by more than increasing the sample size from $n=15000$ to $n=30000$ does (MSE at $n=30000$ of $2131.141$). These results lend credibility to both the proposed method for including larger numbers of replicates and demonstrate that the additional information is useful for improving the quality of the correction.

\begin{figure}
    \caption{\label{fig::simulation_APP2}This plot shows the estimated coverage probability, based on 1000 replicated simulations, for varying levels of nominal coverage using standard, two-sided, normal confidence intervals based on the estimated variance. The dotted blue line indicates a computed coverage equal to nominal coverage.}
    \centering
    \includegraphics[width=0.75\textwidth]{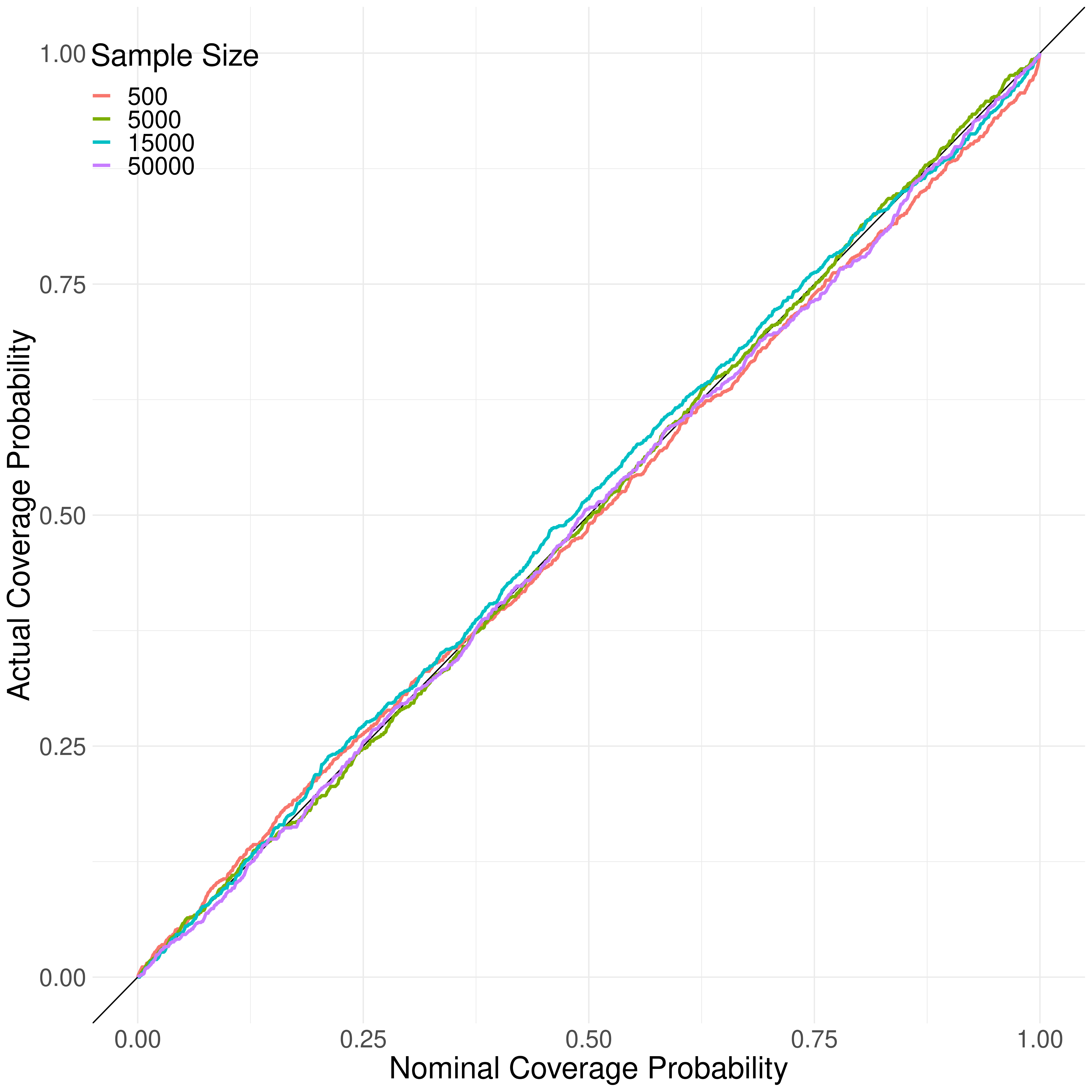}
\end{figure}

Additionally, we investigate the proposed variance estimation technique. Taking the same scenario as in simulation 2, with $n$ to be one of $500$, $5000$, $15000$, or $50000$, we consider using the jackknife inspired variance estimation technique, specifying a quadratic extrapolant for the variance terms. This extrapolant was chosen based on a visual inspection of the plots, rather than through derived theory. The simulations are replicated $1000$ times, and the results are summarized in Figure~\ref{fig::simulation_APP2} and Table~\ref{tab::simulation_APP2}.
\begin{table}
    \caption{\label{tab::simulation_APP2}The actual coverage for selected nominal coverage levels, over the varying sample sizes. }
    \centering
    \begin{tabular}{rrrrr}
      \hline
        Nominal Coverage & $n=500$ & $n=5000$ & $n=15000$ & $n=50000$ \\
        \hline
        0.900 & 0.857 & 0.898 & 0.893 & 0.903 \\
        0.950 & 0.909 & 0.954 & 0.935 & 0.946 \\
        0.990 & 0.969 & 0.995 & 0.990 & 0.990 \\
       \hline
    \end{tabular}
\end{table}
From these results it is clear that this procedure tends to approximate the nominal coverage adequately, supposing that the sample size is sufficiently large. When $n=500$ we see fairly poor coverage results, which tends to improve as $n$ increases. It is worth reiterating that these results assumed a quadratic extrapolant for both the variance estimation and the point estimate. While this quadratic term is theoretically justified for the point estimate, the same justification was +not used for the variance terms. It has been discussed that the quadratic extrapolant tended to be conservative in the standard setting \citep{SIMEX}. While this is generally advisable for a point estimate if in doubt, it is of course less desirable when estimating the variance of an estimator. As a result, higher order extrapolants with less of a tendency to conservatively fit the data may be preferable for this purpose.

This set of simulations considers the use of validation data when the errors are non-symmetric. In particular, we consider the true variate to be distributed according to a Gamma distribution with shape parameter $2$ and scale parameter $1$. The assumed errors have shape parameter $1$ and scale parameter $1.5$. This gives the measurement error a slightly higher variance than the variate itself, a standard deviation ratio of $3/(2\sqrt{2})$. The sample size is taken to be $n=100000$, with a $5\%$ validation sample, that is $m=5000$. The true model for $Y_i$ is a logistic regression, with logit link, with intercept $\beta_0 = 2.5$, slope for $X_i$ as $\beta_1 = -1.25$, and the inclusion of an independent, standard normal variate $Z_i$ with a slope of $\beta_2 = 1$. 

We repeat the simulation $1000$ times, using $B=200$ and $M=10$, and consider the nonlinear extrapolant for all parameters. The results are contained in Table~\ref{tab::simulation_APP3}.

\begin{table}
    \caption{\label{tab::simulation_APP3}The MSE of the estimates of the logistic regression parameters, across $1000$ replicated simulations, comparing a naive fit, the NP-SIMEX, and the standard P-SIMEX procedure. The fit is based on a validation sample of size $5000$, with asymmetric errors.}
    \centering
    \begin{tabular}{lccc}
      \hline
     & Naive & P-SIMEX & NP-SIMEX \\
     \hline
      $\beta_0$ & 0.7631 & 0.2866 & 0.0458 \\
      $\beta_1$ & 0.6608 & 0.1612 & 0.0090 \\
      $\beta_2$ & 0.0415 & 0.0110 & 0.0007 \\
       \hline
    \end{tabular}
\end{table}

\end{document}